\documentclass[aps,prd,twocolumn,nopacs,nokeys,amsmath,amssymb,nofootinbib,preprintnumbers]{revtex4-1}
\usepackage{graphicx}
\usepackage{dcolumn}
\usepackage{amsmath}
\usepackage{bm}
\usepackage{yhmath}
\usepackage{hyperref}
\usepackage{booktabs}
\usepackage{subfigure}
\usepackage{color}
\usepackage[show]{ed}

\usepackage{multirow}
\usepackage{dcolumn}
\newcolumntype{C}[1]{>{\centering\arraybackslash}p{#1}}

\DeclareMathOperator{\Tr}{Tr}

\begin{document}

\preprint{ADP-18-20/T1068}

\title{Accessing high-momentum nucleons with dilute stochastic sources}

\author{J.-J.~Wu}
\author{W.~Kamleh}
\author{D.~B.~Leinweber}
\author{R.~D.~Young}
\author{J.~M.~Zanotti}
\affiliation{CSSM, Department of Physics,
 University of Adelaide, Adelaide SA 5005, Australia}


\begin{abstract}
A novel stochastic technique combining a dilute source grid of $\mathbb{Z}_3$ noise with iterative momentum-smearing is used to study the proton correlation function at rest and in boosted frames on two lattice volumes. The technique makes use of the baryonic version of the so-called one-end trick, and the decomposition into signal and noise terms of the resulting stochastic proton correlation function is made explicit. The number and location of the source points in the dilute grid should be chosen so that the benefits of averaging over many locations overcomes the additional statistical error introduced by the noise terms in the desired fitting region. At all nontrivial momentum values considered we find that the choice of $N=4$--$8$ maximally separated source locations is shown to be optimal, providing a reduced statistical error when compared with a single point source. This enables us to successfully fit the proton energy at momentum values as high as $|\vec{p}| \simeq 3.75$ GeV and $|\vec{p}| \simeq 2.82$ GeV on the small and large volume respectively.
\end{abstract}

\maketitle

\pagenumbering{arabic}

\section{Introduction}\label{sec:I}

%
The study of strong-interaction physics from first principles is expanding in
scope owing to significant advances in lattice QCD technology. The field of
lattice QCD has developed far beyond the study of static observables like masses
and decay constants. 
Modern numerical calculations are now pursuing a range of more
advanced baryonic observables, for example: excited-state
spectroscopy~\cite{Ryan:2015gda,Kiratidis:2015vpa,Briceno:2016mjc,Menadue:2013kfi};
hadron structure~\cite{Lin:2017snn,Chang:2018uxx};
weak transition matrix
elements~\cite{Detmold:2015aaa,Detmold:2016pkz,Sasaki:2012ne,Shanahan:2015dka};
nuclear interactions and electroweak
processes~\cite{Beane:2012vq,Savage:2016kon,Inoue:2011ai}; and the
inclusion of dynamical quantum
electrodynamics~\cite{Borsanyi:2014jba,Horsley:2015eaa}.
This expanded scope of observables brings new challenges to extract quantities
that typically have much weaker statistical signals than the conventional static
observables.
In the present work, we combine the recently developed
momentum-smearing technique~\cite{Bali:2016lva} with a dilute
stochastic grid source~\cite{Gong:2013vja} to improve the isolation of
high-momentum nucleon states in lattice QCD.

Improved correlation functions for accessing hadrons carrying large
momenta have various important applications.
One example is the study of hadron form factors at large momentum
transfer ~\cite{Lin:2010fv,Koponen:2017fvm,Chambers:2017tuf}, where
highly-boosted states are required on one or both sides of the
current.
Recently there has been excellent progress in the numerical study of
partonic structure through
quasi-PDFs~\cite{Lin:2014zya,Alexandrou:2015rja,Chen:2016utp,Alexandrou:2016jqi}
as proposed by Ji~\cite{Ji:2013dva}.
To connect with phenomenological parton distributions, this technique
requires an extrapolation of lattice matrix elements to
$|\,\vec{p}\,|\to\infty$.
Accessing parton distributions directly from the Compton amplitude has
also recently been suggested~\cite{Chambers:2017dov}, where strong
signals are desired at a range of hadronic momenta.
Any resolution of the proton spin puzzle will require lattice QCD
calculations of the gluon spin \cite{Yang:2016plb} contribution as
well as a clear description of the orbital motion of quarks
\cite{Engelhardt:2017miy}.
Both quantities require an extrapolation to
$|\,\vec{p}\,|\to\infty$.

Due to a significant increase in statistical noise at finite
$\vec{p}$, it has been a challenge to reliably study hadron
correlators at large momenta in lattice QCD.
Recently, Bali {\it et al.}~\cite{Bali:2016lva} have demonstrated that
incorporating a momentum phase in the source smearing operation,
preceding quark propagator inversions, can significantly improve the
statistical signal for high-momentum states.
In the present work, we will adapt this technique in combination with
a dilute stochastic source to further improve the statistical signal
at a fixed computational cost.
%

The generation of gauge field configurations requires a significant
amount of computational investment.
It is therefore desirable to gain as much possible information per
gauge configuration.
Because of the finite-ranged nature of QCD, repeated sampling of a
given gauge-field in different spatial (and temporal) locations can
give (almost) independent statistical estimators of hadronic
correlation functions.
In modern simulations, this has seen point-to-all correlators
calculated on as many as 100 or more sites per configuration, with the
statistical scaling being close to the $\mathcal{O}(1/\sqrt{N}$)
expected of independent estimators
~\cite{Beane:2009py,Detmold:2016gpy,Wagman:2016bam}.
Of course the potential gain will depend on explicit factors, such as
the volume, quark mass and observable.
Other than such brute force techniques, innovative techniques have
also been utilised, such as distillation~\cite{Peardon:2009gh},
low-mode averaging~\cite{DeGrand:2004qw}, and
stochastic wall sources~\cite{Gong:2013vja}.

Conventional stochastic wall sources typically lead to very noisy
hadronic correlators.
An exception to this rule would be the so-called ``one end trick'' for
mesons~\cite{McNeile:2006bz, Boyle:2008rh}, which utilises the
conjugation properties of the antiquark.
In general, and particularly for baryons, sampling the source across a
complete set of sites across a 3-volume leads to a large variance
associated with short-distance gauge noise.
Owing to the finite-range correlations of the QCD vacuum, spatially
far-separated points are anticipated to exhibit only a weak
correlation.
A dilute source, sampling multiple sites simultaneously should reduce
the short-distance gauge noise and at the same time achieve
statistical gain by sampling multiple weakly correlated source
locations.
In practice, it should be anticipated that a trade-off is required
where multiple sites increase the signal strength before becoming too
densely packed that the stochastic noise begins to dominate.

To summarise our findings, we find that only a small number of
stochastic sites $\mathcal{O}$(4--8) can be used for each inversion
before the stochastic noise prevents any additional gains.
The increase in statistical precision is found to be more pronounced
for higher momentum states, offering further improvement to the
momentum-smearing technique of Bali {\it et al.}~\cite{Bali:2016lva}.
%

%
In Section~\ref{sec:framework}, we describe our working framework,
including the construction of stochastic sources and our
implementation of momentum-phase smearing.
Our numerical analysis and results are presented in
Section~\ref{sec:analysis}, followed by a summary in
Section~\ref{sec:summary}.
%

\begin{widetext}

\section{Framework}\label{sec:framework}
\subsection{Conventional proton correlation function}
\label{sec:IIa}
%
The standard lattice operator for the proton is
\begin{eqnarray}
\chi(\vec{x}, t) \equiv \epsilon^{abc}\,\left(\, u^{a T}(\vec{x},
t)\, C\gamma_5 d^{b}(\vec{x}, t) \,\right)\,u^{c}(\vec{x}, t),
\label{eq:operator}
\end{eqnarray}
where we are working with Euclidean gamma matrices. This yields
the corresponding two-point correlation function,
\begin{equation}
G(t, \vec{p}, \Gamma) 
\equiv  
 \sum_{\vec{x}}\Gamma\,e^{-i\vec{p}\cdot(\vec{x}-\vec{y})}\,\langle\,
 T\,\left(\, \chi(\vec{x}, t), \bar{\chi}(\vec{y}, 0)\,\right)\,
 \rangle.
\end{equation}
The spatial source position is typically set to the origin,
$\vec{y} \equiv \vec{0},$ but here we allow it to be arbitrary.
The quark propagators $S_f$ for each flavor $u,d$ combine according to
the Wick contractions,
\begin{equation}
 G(t, \vec{p}, \Gamma) = \sum_{\vec{x}}
 e^{i\vec{p}\cdot(\vec{y}-\vec{x})}\,\Gamma^{\gamma\gamma'}\,
 h_{\gamma\gamma'}
 [\,S_u,S_d,S_u\,]\,(\,\vec{x},t;\vec{y},0\,), 
\label{eq:correlationold}
\end{equation}
where the contraction function $h_{\gamma\gamma'}$ is defined as
\begin{multline}
h_{\gamma\gamma'}[S_1, S_2, S_3](\,\vec{x},t;\vec{y},0\,)\equiv 
\epsilon^{abc}\,\epsilon^{a'b'c'}
\,\left\{\,\Tr\left[\,S^{aa'}_{1}(\,\vec{x},t;\vec{y},0\,)\,\gamma_5\,
  C\, S^{bb'\, T}_{2}(\,\vec{x},t;\vec{y},0\,)\,C
  \gamma_5\,\right]\,\left[S^{cc'}_{3}(\,\vec{x},t;\vec{y},0\,)
  \right]_{\gamma\gamma'}\right.\\
\left.
+ \left[\,S^{aa'}_{1}(\,\vec{x},t;\vec{y},0\,)\,\gamma_5 C\, S^{bb'\,
    T}_{2}(\,\vec{x},t;\vec{y},0\,)\,C \gamma_5\,
  S^{cc'}_{3}(\,\vec{x},t;\vec{y},0\,)\,\right]_{\gamma\gamma'} 
\,\right\}.
\end{multline}
\end{widetext}
%
Here, $S_u\,(\vec{x},t;\vec{y},0) =  \langle\,T\left(  u\,(\vec{x},
t), \bar{u}\,(\vec{y}, 0)\,\right)\, \rangle$ is the $u$ quark
propagator, with $S_d$  similarly defined for the $d$ quark.
We assume isospin symmetry $S_u \equiv S_d$ for the proton, and 
the quark flavour index will be dropped henceforth. 

Roman indices $a$--$c\,$($a'$--$c'$) are for colour and Greek indices
$\gamma\,(\gamma')$ are for Dirac spin.
Where it is appropriate, colour and Dirac indices will be implied in
the equations that follow.
The parity projection matrix $\Gamma$ is chosen to be
$\Gamma_+=\left(\, I + \gamma_4 \,\right)/2$.
%

\subsection{Baryon One-End Trick} \label{sec:IIb}

Given a set of spatial noise vectors $\{\xi\}$ with elements drawn
from $\mathbb{Z}_3,$
\begin{eqnarray}
\xi\,(\vec{x}) \in \{ e^{i k\,2\pi/3}; k=0,\pm 1 \},
\label{eq:z3noise}
\end{eqnarray}
then in the (infinite) noise ensemble average we have
\begin{equation}
  \langle \xi(\vec{x}) \xi^\dagger(\vec{y}) \rangle =
  \delta_{\vec{x},\vec{y}},
  \label{eq:deltafn2}
\end{equation}
which is relevant for the meson one-end
trick~\cite{Foster:1998vw,McNeile:2002fh,McNeile:2006bz, Boyle:2008rh,
  Alexandrou:2008ru}. 
For baryons, the required double delta function property is
\begin{equation}
  \langle \xi(\vec{y}) \xi(\vec{y}\,') \xi(\vec{y}\,'') \rangle =
  \delta_{\vec{y},\vec{y}\,'}\delta_{\vec{y}',\vec{y}''},
  \label{eq:z3noisesum}
\end{equation}
which is satisfied for $\mathbb{Z}_3$ noise sources. 

We define a noise source field $\eta$ as the set of $n_{\rm colour}
\times n_{\rm spin}$ fermion vectors with a common spatial dependence
$\xi(\vec{x})$, 
\begin{equation}
  \eta^{aa'}_{\alpha\alpha'}(\vec{x},t) =
  \xi(\vec{x})\,\delta^{aa'}\delta_{\alpha\alpha'}\delta_{t,t_0},
  \label{eq:srcvec}
\end{equation}
where $a,\alpha$ are the fermion indices for colour and spin,
$a',\alpha'$ are the source indices that enumerate the $n_{\rm colour}
\times n_{\rm spin}$ noise vectors, and $t_0$ is the source timeslice.
Then we define for each source vector a corresponding solution vector
\begin{equation}
  \phi^{aa'}_{\alpha\alpha'}(\vec{x},t) = \sum_{\vec{y},b,\beta}
  (M^{-1})^{ab}_{\alpha\beta}(\vec{x},t;\vec{y},t_0)\,
  \eta^{ba'}_{\beta\alpha'}(\vec{y},t_0),
  \label{eq:solnvec}
\end{equation}
where $M$ is the fermion matrix.
\begin{widetext}
Taking the noise ensemble average of the direct product, the stochastic  
estimate of the quark propagator can be written as (suppressing
spin and colour indices),
\begin{equation}
  S(x,y) \simeq  \langle \phi(x)\eta^\dagger(y) \rangle =
  \sum_z
  M^{-1}(x,z) \langle \eta(z)\eta^\dagger(y) \rangle =
  M^{-1}(x,y),
  \label{eq:chitos1}
\end{equation}
where the last equality follows from the delta function property in
equation~(\ref{eq:deltafn2}).

We can generalise the so-called ``one-end trick'' for mesons to baryon
(and baryon-meson) correlators as follows.
Starting from the following single contraction of solution vectors
$\phi$ (repeated indices are summed), we can use equations
(\ref{eq:srcvec}) and (\ref{eq:solnvec}) to expand in terms of the
quark propagator and noise source vectors,
\begin{align} 
(G_2)_{\alpha\gamma'}(t) &= \sum_{\vec{x}}
\epsilon^{abc}\,\epsilon^{a'b'c'}\Gamma_1^{\alpha'\!\beta}
\Gamma_2^{\beta'\!\gamma} \langle
\phi^{aa'}_{\alpha\alpha'}(\vec{x},t)
\phi^{bb'}_{\beta\beta'}(\vec{x},t)
\phi^{cc'}_{\gamma\gamma'}(\vec{x},t) \rangle \label{eq:barphicf}\\
  &= \sum_{\vec{x},\vec{y},\vec{y}',\vec{y}''}
\epsilon^{abc}\,\epsilon^{a'b'c'}\Gamma_1^{\alpha'\!\beta}
\Gamma_2^{\beta'\!\gamma}
S^{ad}_{\alpha\rho}(x,y)S^{be}_{\beta\sigma}(x,y')S^{cf}_{\gamma\tau}(x,y'')\langle
\eta^{da'}_{\rho\alpha'}(y) \eta^{eb'}_{\sigma\beta'}(y')
\eta^{fc'}_{\tau\gamma'}(y'') \rangle\nonumber
\end{align}
where we have left the spin indices $\alpha,\gamma'$ open, $y = (
\vec{y}, t_0 ), y' = ( \vec{y}\,', t_0 ), y'' = ( \vec{y}\,'', t_0 ),$
and $\Gamma_1,\Gamma_2$ are arbitrary spinor matrices appearing in the
interpolating operator of the baryon of interest,
e.g. $\Gamma_1=C\,\gamma_5,\Gamma_2=\gamma_5\,C$ for the proton.
Expanding out the spin and colour dilution indices allows us to apply
the double delta function property (\ref{eq:z3noisesum}) when the
average over $\mathbb{Z}_3$ noise vectors is taken,
\begin{align} 
(G_2)_{\alpha\gamma'}(t) &= \sum_{\vec{x},\vec{y},\vec{y}',\vec{y}''}
  \epsilon^{abc}\,\epsilon^{a'b'c'}\Gamma_1^{\alpha'\!\beta}
  \Gamma_2^{\beta'\!\gamma}
  S^{ad}_{\alpha\rho}(x,y) S^{be}_{\beta\sigma}(x,y')
  S^{cf}_{\gamma\tau}(x,y'') \delta^{da'}
  \delta^{eb'} \delta^{fc'} \delta_{\rho\alpha'} \delta_{\sigma\beta'} 
  \delta_{\tau\gamma'}\langle \xi(\vec{y})\xi(\vec{y}\,')
  \xi(\vec{y}\,'') \rangle \nonumber\\
  &= \sum_{\vec{x},\vec{y},\vec{y}',\vec{y}''}
  \epsilon^{abc}\,\epsilon^{a'b'c'}\Gamma_1^{\alpha'\!\beta}
  \Gamma_2^{\beta'\!\gamma}
  S^{aa'}_{\alpha\alpha'}(x,y) S^{bb'}_{\beta\beta'}(x,y')
  S^{cc'}_{\gamma\gamma'}(x,y'') \delta_{\vec{y},\vec{y}'}
  \delta_{\vec{y}',\vec{y}''}
  \nonumber\\
  &= \sum_{\vec{x},\vec{y}}
  \epsilon^{abc}\,\epsilon^{a'b'c'}\Gamma_1^{\alpha'\!\beta}
  \Gamma_2^{\beta'\!\gamma}
  S^{aa'}_{\alpha\alpha'}(x,y) S^{bb'}_{\beta\beta'}(x,y)
  S^{cc'}_{\gamma\gamma'}(x,y),
  \label{eq:srcsum}
\end{align}
\end{widetext}
demonstrating that after contracting $G_2(t)$ with a spin projection
matrix and taking the trace, e.g. $\mathrm{Tr}\,[\Gamma_+ \,G_2]$ yields
the baryonic two-point correlation function summed over source and
sink positions.
We refer to the set of solution vectors $\phi$ as a stochastic
propagator, noting that they have an identical index structure to a
quark propagator $S,$ so that we can write the standard zero-momentum,
point-source nucleon correlator as
\begin{equation}
G(t, \vec{0}, \Gamma) =\frac{1}{N}\sum_{\vec{x}}
\Gamma^{\gamma\gamma'} \langle\!\langle
h_{\gamma\gamma'}[\phi,\phi,\phi](\vec{x},t)\rangle\!\rangle, 
\end{equation}
where the double angle brackets on the right hand side indicate that
we take the gauge field ensemble average and noise vector ensemble
average concurrently.
Here we also introduce the spatial volume factor $N$, which is needed
to appropriately normalise the sum over source positions $\vec{y}$ in
equation~(\ref{eq:srcsum}).
%

While it is possible to use a noise source $\xi(\vec{x})$ that has
support across the full spatial volume, the resulting stochastic
estimate of the double delta function is very noisy, to the point that
the signal for the ground state nucleon is washed away by the
statistical fluctuations.
This statistical noise comes from the cross-terms between different
grid points when the product of noise vectors is expanded, and are
suppressed at large spatial separation.
We note that the above formalism for the baryon one-end trick also
holds true for spatially diluted noise sources, where we
systematically set $\xi(\vec{x}) = 0$ on some subset of the spatial
volume.
An all-to-all calculation of the two-point correlator can be achieved
by inverting across multiple diluted sources and summing the results,
but this requires a significant increase in the matrix inversion
count.

The alternative that is investigated here is the use of a single
highly dilute noise source, such that in the solution
field~(\ref{eq:solnvec}), we can restrict the sum over $\vec{y}$
to the $N$ non-vanishing grid sites $\vec{y}_n$ of the dilute source,
\begin{equation}
\phi(\vec{x},t) =\sum_{n=1}^{N}
S(\vec{x},t;\vec{y}_n,0)\,\eta(\vec{y}_n,0).
\label{eq:stochi1}
\end{equation}
In subsequent equations it is to be understood that $\eta$ is only
nonvanishing on a subspace of a fixed-$t$ wall. In practice, $N$ will
be much less than the lattice 3-volume.
This is motivated by observing that inverting from a single noise
source on a dilute grid allows us to average over a small number $N$
of source points that are at large spatial separation so as to
minimise the statistical noise, potentially providing an advantage
when compared to a single point source.
\begin{figure}[tpb]
\begin{center}
\includegraphics[width=0.45\textwidth]{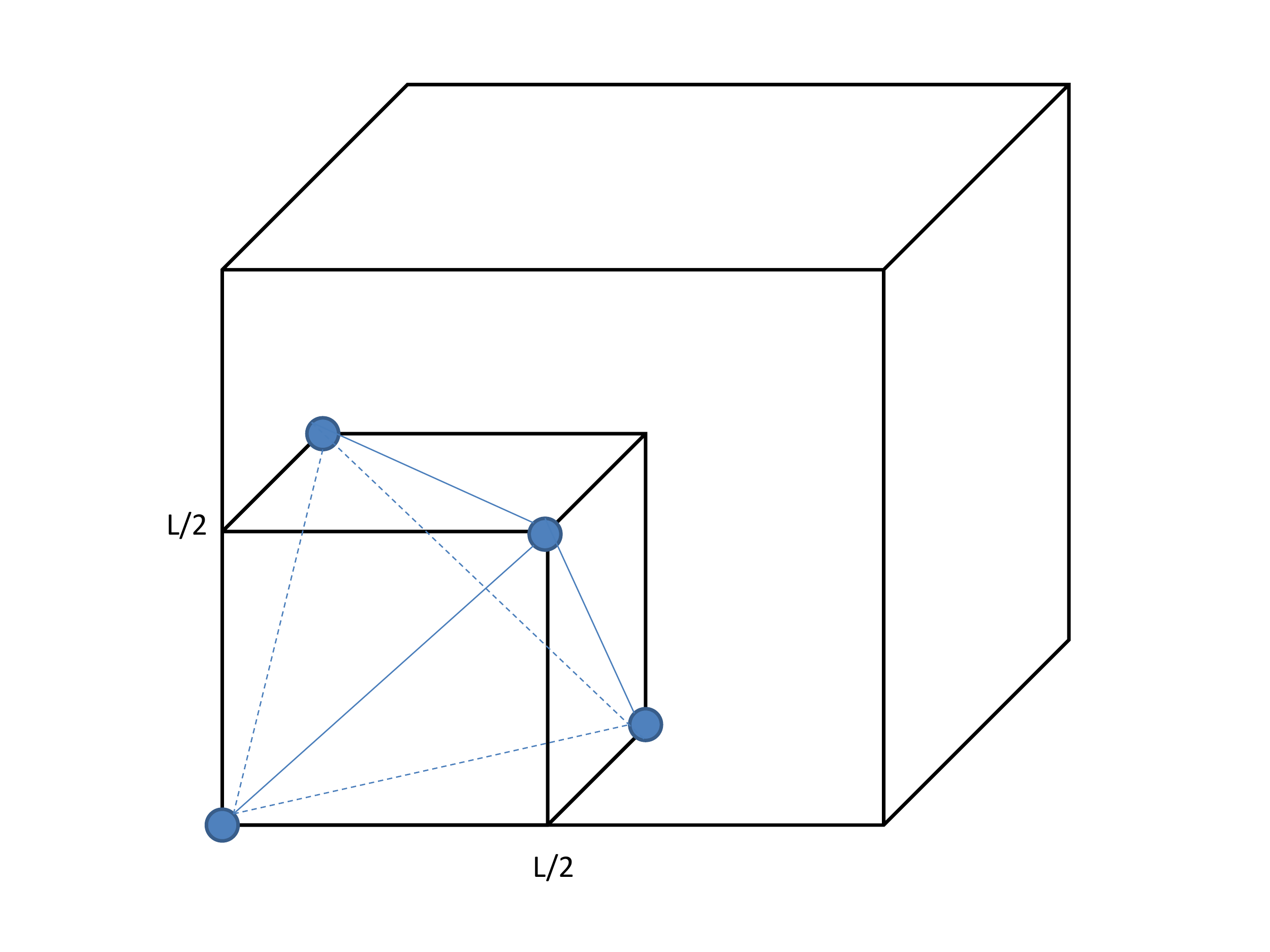}
\caption{ The $N=4$ source points are located on the four vertics of a
  tetrahedron with edge length $L/\sqrt{2}$.}
\label{fg:dilute}
\end{center} 
\end{figure}

Figure~\ref{fg:dilute} shows the choice of non-vanishing sites for
$N=4$, which maximises the distance of any pair of the points
considered.

\subsection{Quark momentum phase in the source}
\label{sec:IIc}
\subsubsection{Fourier phase}
To study states at finite momentum, one must correlate the phase
appropriately between the source and sink location.
It is clear from Eq.~(\ref{eq:barphicf}) that when contracting
the solution vectors, only the sink location $\vec{x}$ is directly
accesible.
As the stochastic propagator $\phi$ includes various source locations,
in order to coherently project the nucleon to a non-zero momentum
$\vec{p}$ the appropriate Fourier phase for the momentum $\vec{q}$
carried by each quark must be applied to each noise source point
before the fermion matrix inversions are performed,
\begin{equation}
  \xi_{\vec{q}}(\vec{y}) = e^{i\vec{q}\cdot\vec{y}}\xi(\vec{y}). \label{eq:phaseFourier}
\end{equation}

Applying equation~(\ref{eq:stochi1}) then yields a stochastic
propagator $\phi_{\vec{q}}$ that implicitly encodes the appropriate
Fourier phase for a quark with momentum $\vec{q}$ at each source
location, such that the nucleon correlator at a specific momentum
$\vec{p}$ can be obtained by applying the standard Fourier projection
at the sink location,
\begin{equation}
G(t, \vec{p}, \Gamma) =\frac{1}{N}\sum_{\vec{x}}
e^{-i\vec{p}\cdot\vec{x}} \Gamma^{\gamma\gamma'} \langle\!\langle
h_{\gamma\gamma'} [\phi_{\vec{q}_1},\phi_{\vec{q}_2},\phi_{\vec{q}_3}]
(\vec{x},t)\rangle\!\rangle,
\label{eq:correlation}
\end{equation}
To ensure the coherent signal from all source locations, the sum of
the quark Fourier momenta must equal the total momentum of the hadron,
$\sum \vec{q}_i=\vec{p}$.
With this condition, it is straightforward to expand the above in the same manner as
Eq.~(\ref{eq:barphicf}) to show that the resulting nucleon
correlator acquires the appropriate Fourier phase of
$e^{-i\vec{p}\cdot(\vec{x}-\vec{y})}$ for each source location
$\vec{y}.$

\subsubsection{Smearing phase}

We apply iterative momentum smearing~\cite{Bali:2016lva} to our
lattice operators to improve the signal at high momentum.
Iterative momentum smearing modifies the spatial links in the standard Jacobi
smearing procedure \cite{Best:1997qp} to include a momentum factor $\vec{k}$,
such that the fermion source after $m+1$ smearing sweeps is given by
\begin{align}
  \eta_{m+1}(x) &= \eta_0(x) +\rho\sum_{j=1}^3 \bigg[e^{i\vec{k}\cdot\hat{e}_j}U_j(x)\,\eta_m(x+\hat{e}_j)\nonumber\\
    &\quad\quad\quad+e^{-i\vec{k}\cdot\hat{e}_j}U_j^\dagger(x-\hat{e}_j)\,\eta_m(x-\hat{e}_j)\bigg].
\end{align}
In this work we construct the momentum smeared propagator $S_{\vec{k}}$ by
applying 60 sweeps of iterative momentum smearing at the source and the sink
with a Jacobi smearing factor $\rho = 0.21$.

The composite solution vector to the quark matrix inversion now becomes
\begin{equation}
\phi_{\vec{q},\vec{k}}(\vec{x}, t) = \sum_{n=1}^{N}  e^{i\vec{q}\cdot
  \vec{y}_n} S_{\vec{k}}(\vec{x},t;\vec{y}_n,0)\,\eta_0(\vec{y}_n,0).
\label{eq:propaq}
\end{equation}
In principle, the source smearing momenta can be chosen arbitrarily~\cite{Bali:2016lva}, with the values being optimised to improve the overlap of the operator with a hadron of chosen momenta.
In the numerical calculations reported this work, the source/sink smearing momentum $\vec{k}$ is set to be equal to the corresponding quark Fourier momentum $\vec{q}$ of Eq.~\ref{eq:phaseFourier}.

\subsection{Proton correlation function with momentum-based noise source}
\label{sec:IId}

%
The correlation function of the proton is computed using the
stochastic propagator $\phi_{\vec{q},\vec{q}}(\vec{x},t)$, instead of
$S(\vec{x},t;\vec{y},0)$.
Correspondingly, we only need to keep track of the Fourier phase at
the sink, as the source phase has already been absorbed into the
stochastic propagators.
Letting $\vec{p} = \vec{q}_1+\vec{q}_2+\vec{q}_3$ be the sum of the quark Fourier
momenta, then using equation~(\ref{eq:propaq}), along with the
$\mathbb{Z}_3$ noise property $\eta^3(\vec{y}_n,0) = 1,$ we expand the
proton correlation function (\ref{eq:correlation}) into two parts,
\begin{widetext}
\begin{multline}
\label{eq:correlationd}
G(t, \vec{p}, \Gamma) = \frac{1}{N}\sum_{\vec{x}}
\Gamma^{\gamma\gamma'} \Bigg\{\sum_n
e^{-i\vec{p}\cdot(\vec{x}-\vec{y}_n)}\langle\!\langle
h_{\gamma\gamma'} [S_{\vec{k}_1},S_{\vec{k}_2},S_{\vec{k}_3}]
(\vec{x},t;\vec{y}_n,0)\rangle\!\rangle\,+ \\
\sum_{n,l,m}e^{-i(\vec{p}\cdot\vec{x} - \vec{q}_1\cdot\vec{y}_n -
  \vec{q}_2\cdot\vec{y}_l - \vec{q}_3\cdot\vec{y}_m)}
(1-\delta_{nl}\,\delta_{nm}) \langle\!\langle h_{\gamma\gamma'}
    [S_{\vec{k}_1},S_{\vec{k}_2},S_{\vec{k}_3}]
    (\vec{x},t;\vec{y}_{[n;l;m]},0)\, \eta(\vec{y}_n,0)\,
    \eta(\vec{y}_l,0)\,\eta(\vec{y}_m,0)
    \rangle\!\rangle\,\Bigg\},\\
\end{multline}
\end{widetext}
where the notation $\vec{y}_{[n;l;m]}$ implies that the different
source location indices $n,l,m$ are paired with the appropriate
propagator $S_{\vec{k}_1},S_{\vec{k}_2},S_{\vec{k}_3}$ in the
expansion of the contraction function $h_{\gamma\gamma'}.$
The first part is simply the summation over the $N$ source points of
the standard proton correlation function in
Eq.~(\ref{eq:correlationold}) (using the smeared propagator), and
hence we refer to these as the signal terms in the following
discussion.
For uncorrelated spatial source point separations, the error of the
signal terms should be smaller than Eq.~(\ref{eq:correlationold}) by a
factor of $1/\sqrt{N}.$
On the other hand, the second part of equation~(\ref{eq:correlationd})
will go to zero when averaged across a large number of noise sources
because of the double delta function property~(\ref{eq:z3noisesum}).
We refer to the terms in the second part as noise terms, as they
contain the product of noise vectors at distinct locations (which
should vanish), and as such are a new source of statistical error in
the stochastic proton correlation function~(\ref{eq:correlation}).
Clearly, to get a better signal, we should make the signal terms
stronger and suppress the noise terms, and later we will show how to
choose the $N$ source locations toward this aim.
%

%
We can maximise the accessible proton momenta values by judiciously
choosing the set of three-momenta used to calculate each quark
propagator.
Here, we calculate quark propagators with four different values of the
three-momenta at the source,
\[
\vec{q}=\{ (0,0,0),(0,0,1),(0,1,1),(1,1,1)\},
\]
enabling us to generate 20 distinct total proton momenta from
$(0,0,0)$ to $(3,3,3)$.
%

%
 
\section{Results}\label{sec:analysis}

We study the proton using the above dilute noise source on dynamical
lattices generated with the Wilson gluon action and $N_f=2$ flavours
of nonperturbatively improved Wilson femions.
Two lattice volumes are used, $24^3\times48$ (376 configurations), and
$32^3\times64$ (1000 configurations), both at a gauge coupling of
$\beta = 5.29,$ corresponding to an inverse lattice spacing of $a^{-1}
\sim 2.76\text{ GeV}$.
The hopping parameters are $\kappa=0.1355$ and $0.13632$, providing
pion masses of $m_\pi = 902$ GeV and $m_\pi = 295$ GeV for the smaller
and larger volume, respectively.
In the following we will often use $\vec{P}$ to refer to the triplet
of integers specifying the momenta in lattice units, and use $\vec{p}
= (2\pi/L)\vec{P}$ to refer to the physical momenta.

\subsection{Stochastic error terms}
\label{sec:IIIa}

%
The stochastic estimation of the double delta function
(\ref{eq:z3noisesum}) is a new source of error in the proton
correlation function, encapsulated by the noise terms in
Eq.~(\ref{eq:correlationd}).
Thus, it is desirable to find ways to minimize these terms.
%

%
We denote by $G_N$ the proton correlator obtained from stochastic
propagators with a dilute grid of $N$ non-zero source points, and
define $\sigma_N$ to be the corresponding statistical error in the
full ensemble average over gauge fields and noise vectors.
For the case $N=2$ we can isolate the correlation function from the
signal terms by calculating an independent point-to-all correlator for
each of the two source locations $\vec{y}_1,\vec{y}_2.$
This can be compared with the correlation function obtained from the
stochastic propagators, which combines the signal and noise terms.

\begin{figure}[tb]
\begin{center}
\includegraphics[width=0.5\textwidth]{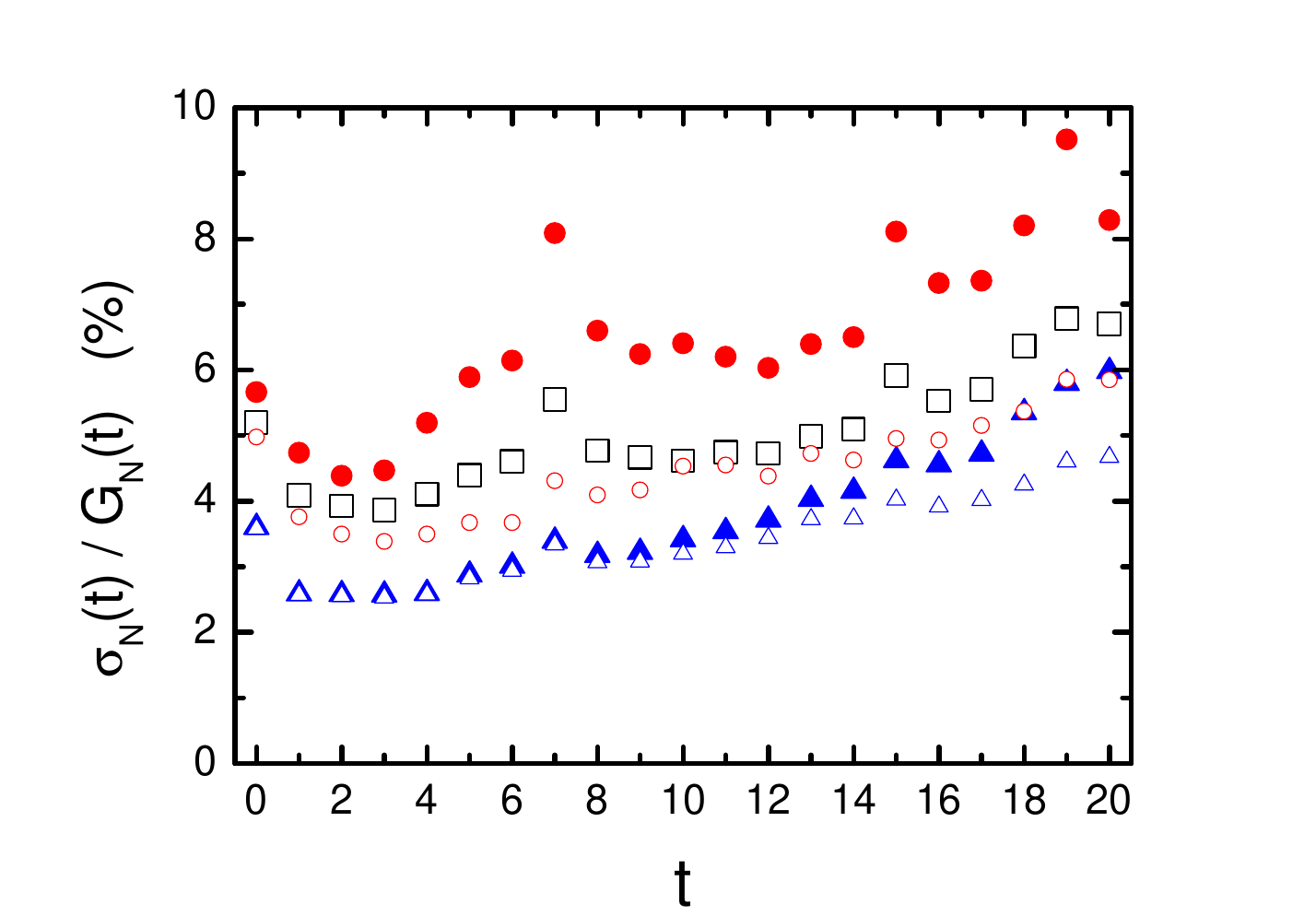}
\caption{
%
The relative error in the proton correlation function on the
$24^3\times 48$ lattice with $\vec{p}=(0,0,0)$.
The open square points use a single-source location $\vec{y} =
(0,0,0)$, while the circle and triangle points are calculated using
two source points,  $\vec{y}_k = (0,0,0), (2,2,2)$ and
$\vec{y}_k=(0,0,0), (12,12,12)$, respectively.
The open points represent the signal terms in
Eq.~(\ref{eq:correlationd}), while the solid points include both the
signal and noise terms.
%
}\label{fg:twopoints}
\end{center} 
\end{figure}

\begin{figure}[tb]
\begin{center}
\includegraphics[width=0.5\textwidth]{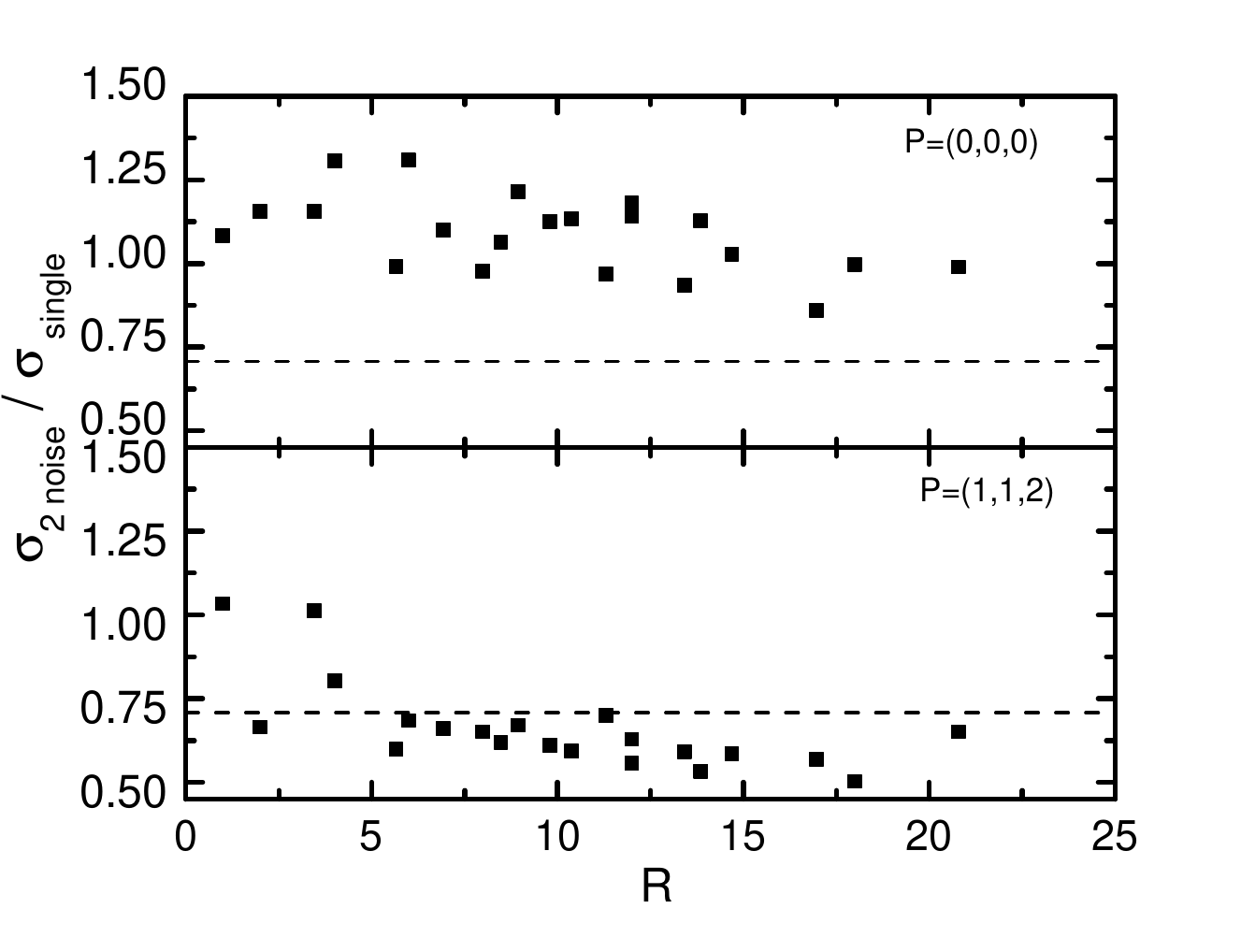}
\caption{ 
The statistical error in the fitted energy of the proton as a
function of the spatial separation $R$ (in lattice units) between
source locations for a dilute grid with $N=2$ points.
Results are shown for the rest frame $\vec{P}=(0,0,0)$ (top), and a
boosted frame with $\vec{P}=(1,1,2)$ (bottom) on the $24^3\times 48$
lattice.
The dashed lines indicate the expected relative improvement of $1/\sqrt{2}$ in the error for having two independent sources.
%
}\label{fg:twopoints2}
\end{center} 
\end{figure}

In Fig.~\ref{fg:twopoints}, we show the relative error $\sigma_N/G_N$
as a function of Euclidean time for various spatial source grids with
$N=1,2.$
The relative errors of a correlation function calculated from a
single-source location $N=1$ (black open squares) are comparable or
slightly larger than the signal terms arising from a source with
closely spaced locations $\vec{y}_1=(0,0,0)$ and $\vec{y}_2=(2,2,2)$
(red open circles).
However, after including the noise terms (red solid circles), the
errors increase quickly and grow larger than those from a single point
source.
This enhancement of errors is purely due to the noise term
contributions from source points that are close together.
When we change the two source locations to be further apart at
$\vec{y}_1(0,0,0)$ and $\vec{y}_2=(12,12,12)$ (blue triangles), it is
clear that the errors both with and without noise terms become
similar, and are much smaller than the above two cases (circles and
squares).
This demonstrates the value of having maximally separated source
locations.

In Fig.~\ref{fg:twopoints2}, we show the ratio of the absolute errors
in the fitted proton energy for $N=2$ source locations compared to
a single source location, as function of the distance between the
$N=2$ source locations.
We find that the absolute error in the energy decreases as
the distance between two source locations increases, indicating that
the two sources are becoming decorrelated at larger separations.
It is interesting to note that the rate at which the error decreases
with separation is momentum dependent.
In the rest frame $\vec{P}=(0,0,0),$ the reduction is slow, and we do
not achieve the ideal improvement factor of $1/\sqrt{2}$ for any of
the source separations studied.
However, for a large momentum boost $\vec{P}=(1,1,2),$ the error drops
rapidly, then plateaus at or below the ideal $1/\sqrt{2}$ dashed
line.
Note that any values below the ideal line are an artefact of a finite
statistical ensemble, as in the limit of infinite separation the
source locations are independent.

From the above comparisons, we see that the distance between source
locations will play an important role in minimizing noise terms.
This is expected, as the quark propagator suppresses the size of the
noise terms by the distance between the source points.
If the source locations are sufficiently spaced the contribution from
noise terms becomes negligible.
%
  
%
To keep the source points appropriately spaced, the number of source
locations, $N$, should not be very large.
As shown in Eq.~(\ref{eq:correlationd}), the number of noise terms is
$\mathcal{O}(N^3)$, while the number of signal terms is
$\mathcal{O}(N)$.
Adding additional source points provides more averaging, but also
decreases the maximal source separation. To balance these two
competing effects, we investigate if there exists an optimal choice
for each value of $N$.

%
%
%
%
%

\begin{figure}[tbp]
\begin{center}
\includegraphics[width=0.5\textwidth]{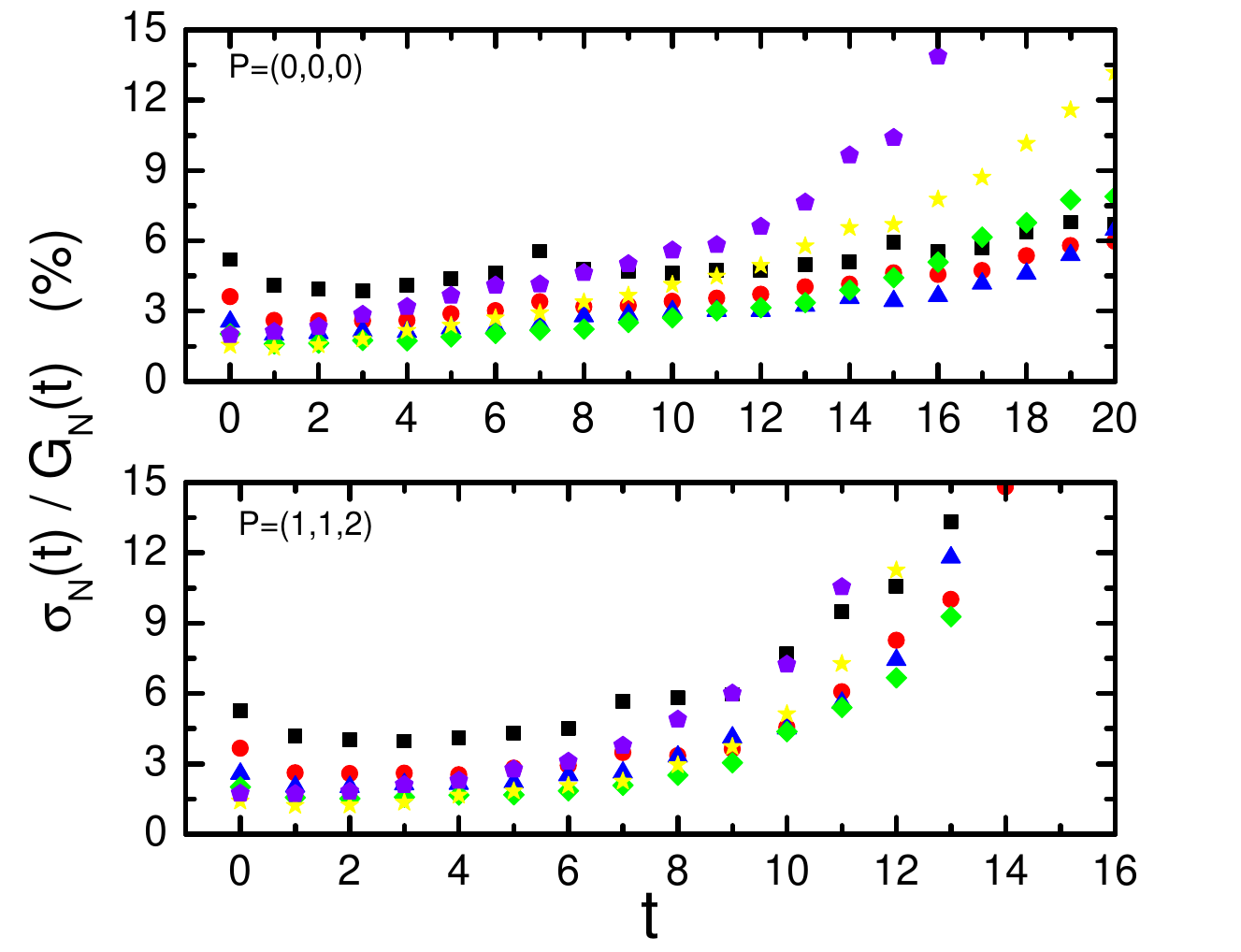}
\caption{
%
The relative error in the proton correlation function on the
$24^3\times 48$ lattice in the rest frame (top) and a boosted frame
with $\vec{P}=(1,1,2)$ (bottom).
The black square, red circle, blue triangle, green diamond, gold star,
and purple pentagon points are calculated from $N=1$, $2$, $4$, $8$,
$27$ and $64$ source locations respectively.
The $N=2$ source locations are $(0,0,0)$ and $(12, 12, 12)$.
The $N=4$ locations are the four vertices of a tetrahedron with edge
length $12\sqrt{2}$.
The $N=8, 27$ and $64$ locations are chosen to lie on a cubic grid,
separated by 12, 8 and 6 lattice spacings in each spatial direction.}
\label{fg:Npoints}
\end{center} 
\end{figure}

\begin{figure}[tbp]
\begin{center}
\includegraphics[width=0.5\textwidth]{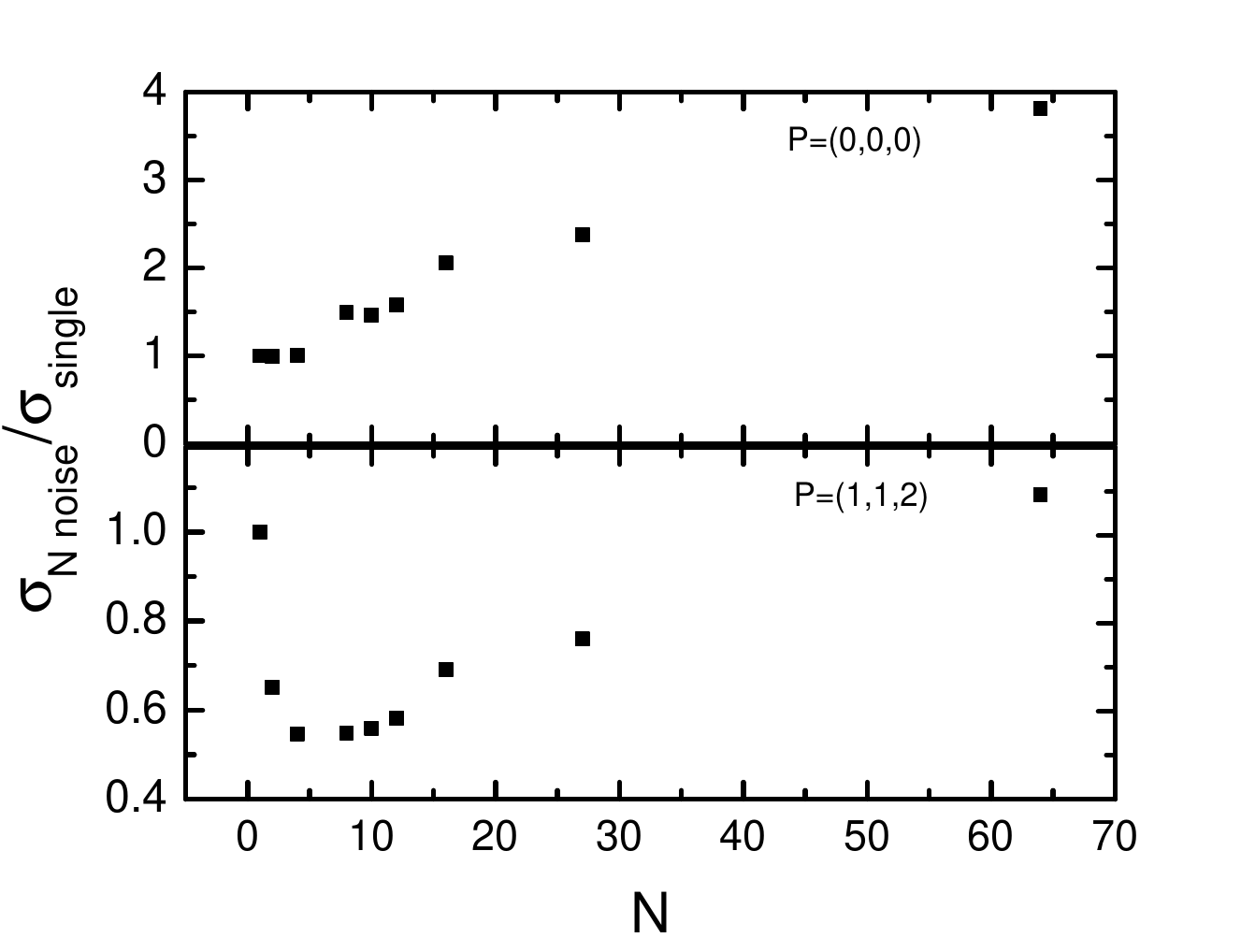}
\caption{ The error in the effective energy of the proton with total
  momentum $\vec{P}=(0,0,0)$ (top) and $(1,1,2)$ (bottom).
  $N$ is the number of source locations.
  For each value of $N$, the source locations are chosen to maximise
  their separation (as described in Fig.~\ref{fg:Npoints}).}
\label{fg:Npoints2}
\end{center} 
\end{figure}

In Fig.~\ref{fg:Npoints}, the relative error $\sigma_N/G_N$ for
different choices of source location are shown for zero momentum (top)
and at $\vec{P}=(1,1,2)$ (bottom).
Maximally separated source locations are chosen for each value of
$N$.
At non-zero momentum, we find that $N=4$ (triangles) and $N=8$
(diamonds) provide smaller errors than the other values of $N.$
The errors in the proton effective energy as a function of the number
of source points are shown in Fig.~\ref{fg:Npoints2}, for the same two
choices of $\vec{P}=(0,0,0)$ (top) and $\vec{P}=(1,1,2)$ (bottom).
Once again, the source locations are maximally separated for each $N.$
In the upper plot at zero momentum, the smaller values of $N=1,2,4$
are best with a similar error, while in the lower plot at
$\vec{P}=(1,1,2)$, we see that as $N$ increases the error drops
rapidly at first and then increases quickly, with $N=4, 8$ providing
the best results.

%
In summary, the choice of source locations will directly affect the
quality of the signal.
It is important that the source points are chosen to be sufficiently
spaced in order to suppress the contribution from the noise terms.
Overall, our results for the $24^3\times48$ lattice suggest that the
best choice for the number of source locations is around $N=4 - 8,$
and we will now proceed to study the proton correlation function at a
large number of momenta on this volume.
%

\subsection{$24^{3}\times48$ results }
\label{sec:IIIb}

\begin{figure}[tbp]
\begin{center}
\includegraphics[width=0.46\textwidth]{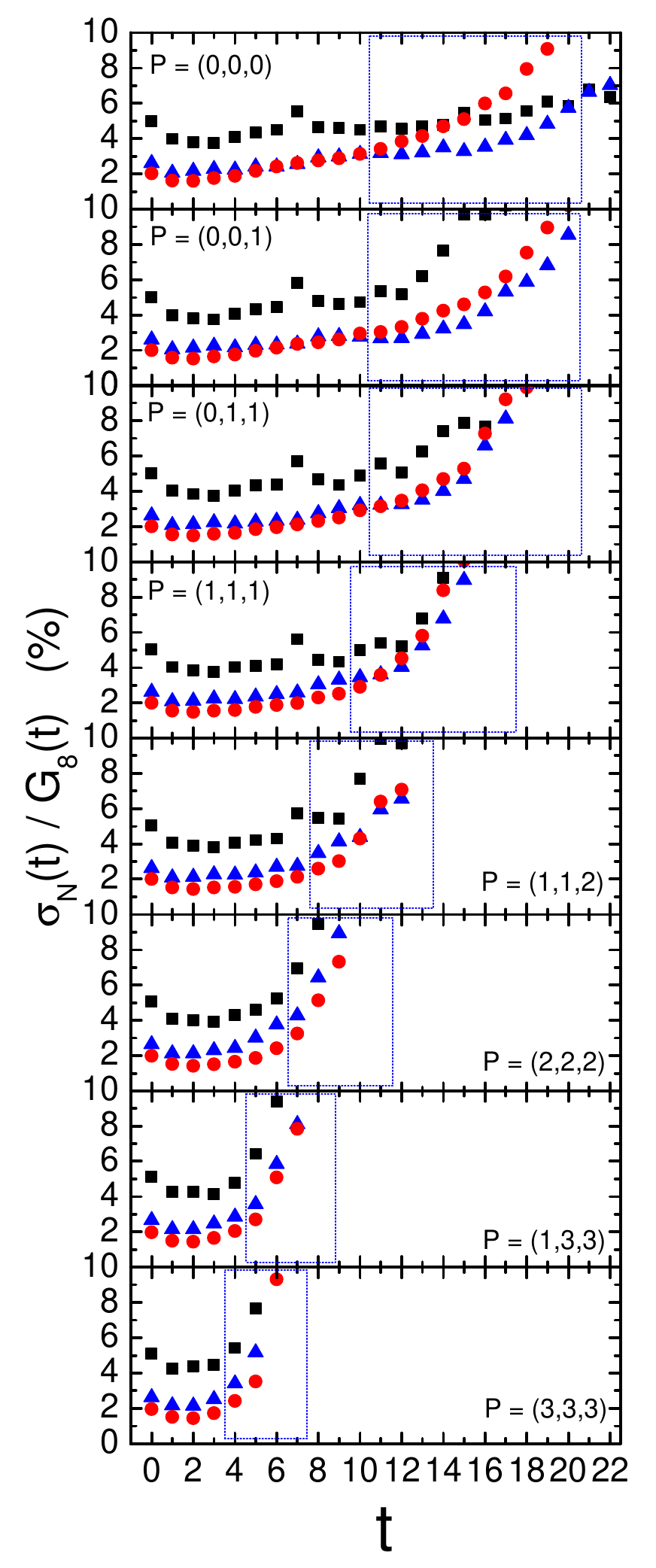}
\caption{ The relative error in the proton correlation function with
  $N$ source points for the $24^3\times 48$ lattice volume.
The black square, blue triangle, and red circle points are for $N=1$,
$4$ and $8$ source locations, respectively.
The dashed blue boxes show the selected window for fitting the
effective energy.}
\label{fg:correlationL24}
\end{center} 
\end{figure}

\begin{figure}[tbp]
\begin{center}
\includegraphics[width=0.45\textwidth]{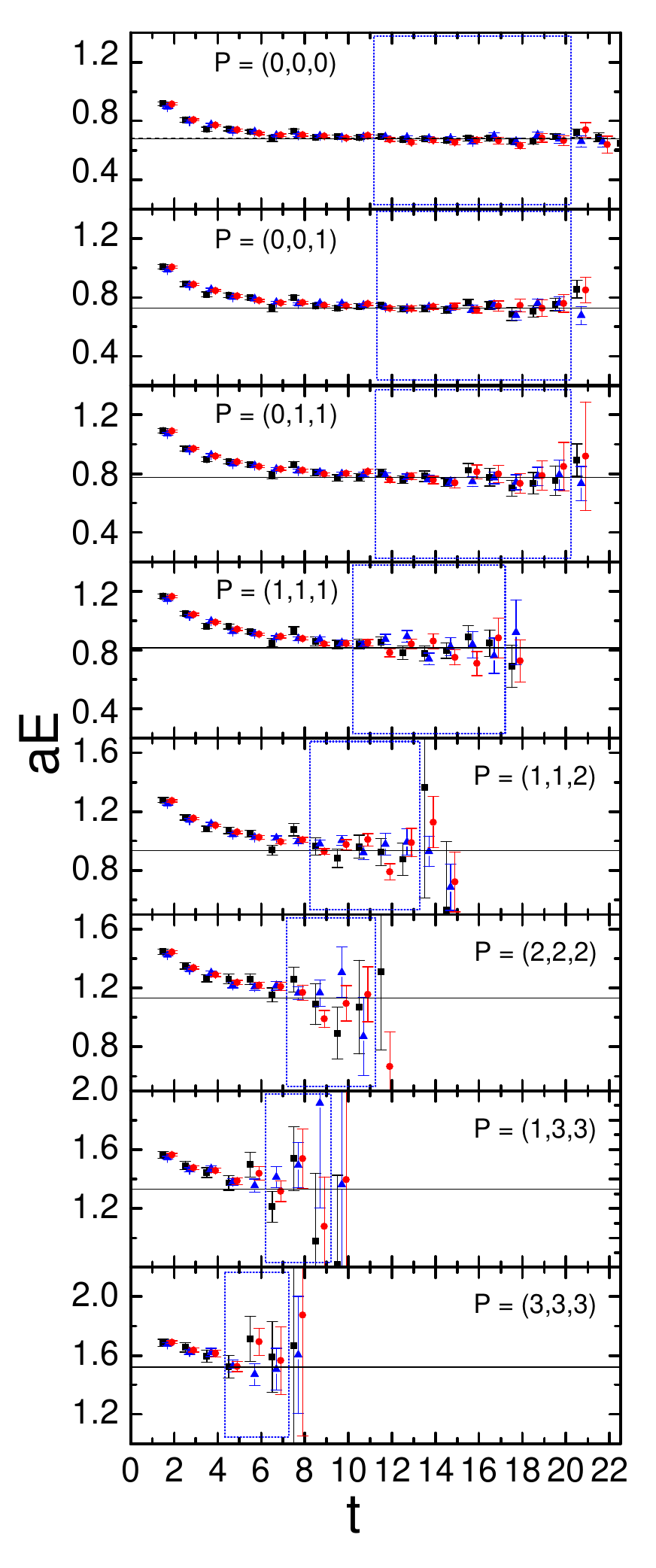}
\caption{ 
The proton effective energy in lattice units on the $24^3\times 48$
volume, with an inverse lattice spacing of $a=2.76\text{ GeV}^{-1}.$
Values are shown at eight momenta for $N=1,4,8$ with the same symbols
as in Fig.~\ref{fg:correlationL24}.
The solid line in the rest frame is the effective mass of the proton,
$am$, calculated from the average of the three fitted effective masses
of the proton from the values of $N$. 
The solid lines at nonzero momenta are calculated from the continuum
dispersion relationship, $ aE = a\sqrt{m^2 + p^2}$.
The blue dashed boxes show the fitting range.}
\label{fg:amdataL24}
\end{center} 
\end{figure}

\begin{figure}[btp]
\begin{center}
\includegraphics[width=0.45\textwidth]{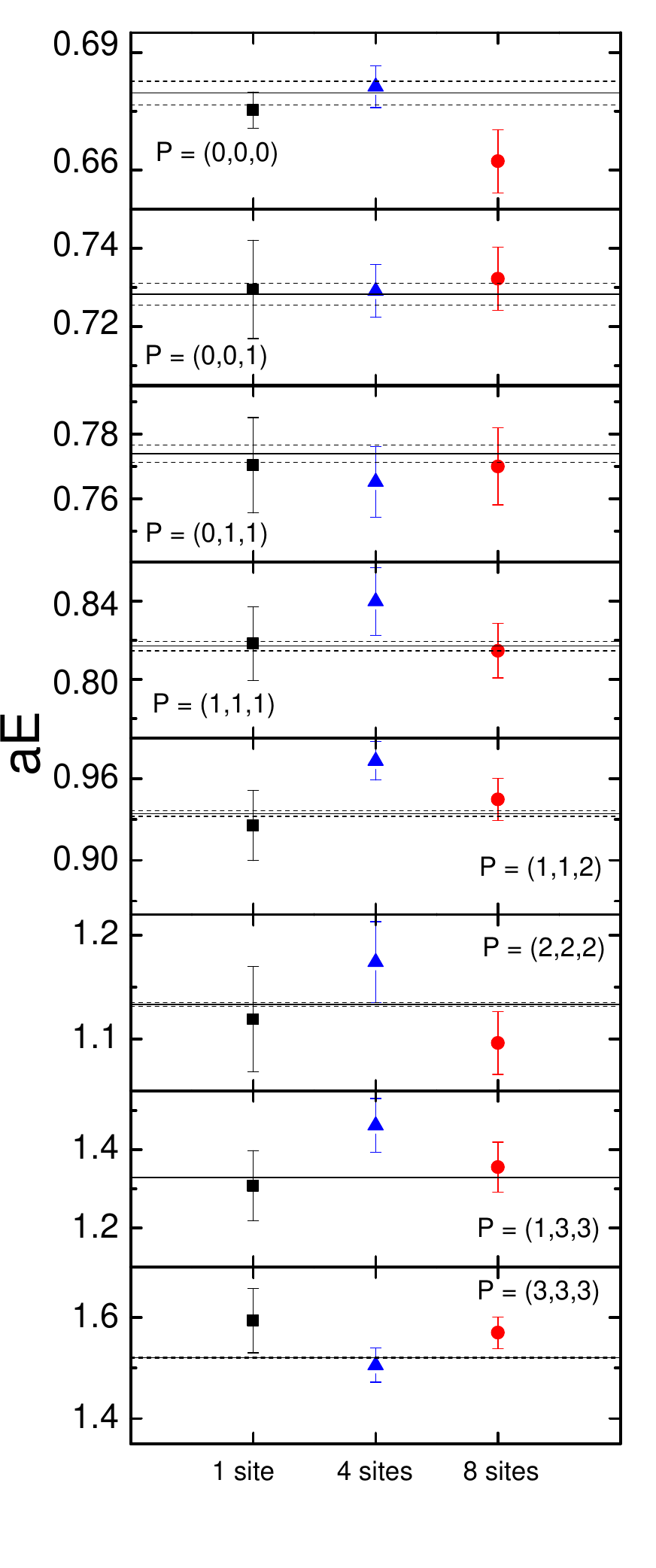}
\caption{ The fitted proton energies in lattice units for the
  $24^3\times 48$ volume.
  Values are shown at eight momenta for $N=1,4,8$ with the same
  symbols as in Fig.~\ref{fg:correlationL24}.
The solid lines shows the central value for the continuum dispersion
relationship, $ aE = a\sqrt{m^2 +  p^2},$ with the dashed lines
indicating the corresponding estimate of the error derived from the
fitted proton mass $m$.
We note that at all non-trivial momentum values the statistical errors
for $N=4,8$ source locations are reduced compared to the point source
$N=1$.}
\label{fg:amfitL24}
\end{center} 
\end{figure}

%
The proton correlation function on the $24^{3}\times48$ lattice is
studied by analysing the relative error and effective energy at eight
different momentum values $\vec{P} \in \{ (0,0,0), (0,0,1), (0,1,1),$
$(1,1,1), (1,1,2), (2,2,2),$ $(1,3,3), (3,3,3) \},$ with results shown
for $N=1$ (black square), 4, (blue triangle) and 8 (red circle) source
locations in Figures~\ref{fg:correlationL24}-~\ref{fg:amfitL24}.
To separate the source locations as much as possible, for $N=4$ the
four vertices of a tetrahedron with edge length $12\sqrt{2}$ are
chosen, while for $N=8$ the points are the eight vertices of a cube
with edge length $12$ (in lattice units).
%

The statistical error in the proton correlation function for the three
different sources at each of the eight momenta are shown in
Fig.~\ref{fg:correlationL24}.
To compare the errors across different values of $N$, the relative
error is constructed with respect to the central value of the
correlation function for $N=8,$ since overall that provides the best
signal here.
Firstly, we note that at small times $t$, the error for $N=8$ is much
smaller than that of a point source, by at least a factor of 2.
However, as Euclidean time increases, the error for all three cases
increases, and the rate of increase is faster for larger $N.$
This is understood by noting that in the correlation function the
noise terms originating from the different source locations will grow
larger with time evolution, as the extent of the wave function of the
quark expands.
Thus, using a dilute noise grid provides a much more precise signal at
early times.

The blue boxes in Figs.~\ref{fg:correlationL24} and \ref{fg:amdataL24} indicate the selected fitting ranges of 
the effective energy illustrated in Fig.~\ref{fg:amdataL24} for each momentum.
In these fitting ranges, with the exception of the rest frame, using
four and eight source locations gives a better signal than a single
source location for each non-trivial momentum value.
For the rest frame, we observe that the error for 8 source points
increases rapidly, such that in the fitting range the error exceeds
the others, while 4 source points remains better than a single source.
Through these comparisons, we find that using an appropriate dilute
$\mathbb{Z}_3$ noise source can provide a significant benefit in
obtaining the effective energy of proton in a boosted frame, relative
to a standard point source.
%

The energies obtained from the fit windows of Fig.~\ref{fg:amdataL24} are displayed in Fig.~\ref{fg:amfitL24} for each proton momentum considered.
As was observed previously in Fig.~\ref{fg:correlationL24}, the
effective mass for $N=4$ and 8 source locations have a small error at
early times, while suffering from larger errors at large times.
Therefore, the important step is to choose a suitable fitting window,
for which we apply the following steps.
First, we fix the latest time $t_{\rm max}$ to fit by considering the
relative error in the data for all three values of $N$, excluding any
points that are too noisy. For example, at $P=(1,1,2)$, we exclude
data above $t=13.5$ because the errors for $N=1$ (black squares) blow
up at this point.
Next, the earliest time $t_{\rm min}$ is fixed by considering the
reduced $\chi^2$ obtained by fitting the correlation function up to
$t_{\rm max}$.
In this case, we require that the reduced $\chi^2$ for all three values of
$N=1,4,8$ are smaller than 1.5, such that the fitting windows coincide
for the purposes of our comparison.
Finally, some care must be applied to avoid fitting before the
correlation function is dominated by a single energy state,
particularly at higher momenta where the signal is noisy.
Hence, we do not consider early time regions where there is a clear
systematic downward drift in the effective energy for our fits.

Examining the selected fitting windows in Figure~\ref{fg:amdataL24},
we see that the first three momentum values share a common window, but
after that the increase in noise at higher momenta values forces us to
move to earlier times. Even at the largest value of
$\vec{P} = (3,3,3)$ we are able to find a plateau before the signal
has degraded.
In Figure~\ref{fg:amdataL24}, each sub-plot uses a common vertical
scale, though the upper and lower bounds vary for each value of
$\vec{P}.$

The resulting fit values and errors are shown in
Figure~\ref{fg:amfitL24} and indicate that using a dilute grid
source with $N=4,8$ gives more accurate energies for the
proton, with significantly reduced errors at all values of $\vec{P},$
with the exception of the rest frame.
Unlike the previous figure, in Fig.~\ref{fg:amfitL24} each sub-plot
uses an independent vertical scale, so that we are able to better
compare the results for the three different values of $N$ at each
momenta.

In the rest frame, the fit error for 4 source locations is comparable
to a single source, but the error for 8 source points is larger.
This is understood by noting that the number of noise terms at $N=4$
is reduced by a factor of $2^3$ relative to $N=8,$ which provides a
better overall signal in the rest frame, as the fit window is large
enough to be affected by the late time behaviour of the error
increasing with $N$ (as seen in Fig.~\ref{fg:correlationL24}).

The relative error in the correlation function for a single source in
the rest frame is around $6\%$ for the selected fit window, increasing
to $10\%$ or greater in a moving frame, even at $\vec{P} = (0,0,1)$.
When we consider the results at finite momentum, we find that for
$N=4,8$ source points the fit error is smaller that that for a single
source (unlike in the rest frame).
This is true even at large times.

The observation that at finite momentum the benefits of the additional
averaging outweigh the errors from the noise term contributions may be
explained by our expectation that the noise term suppression by the
quark propagator for spatially separated sources should increase at
large momentum.
The motivation for this explanation is that the characteristic scale
for the physics of the nucleon will decrease at high momentum, hence
points at a fixed spatial separation will become less correlated at
higher values of $\vec{P}.$
\begin{figure}[tbp]
\begin{center}
\includegraphics[width=0.55\textwidth]{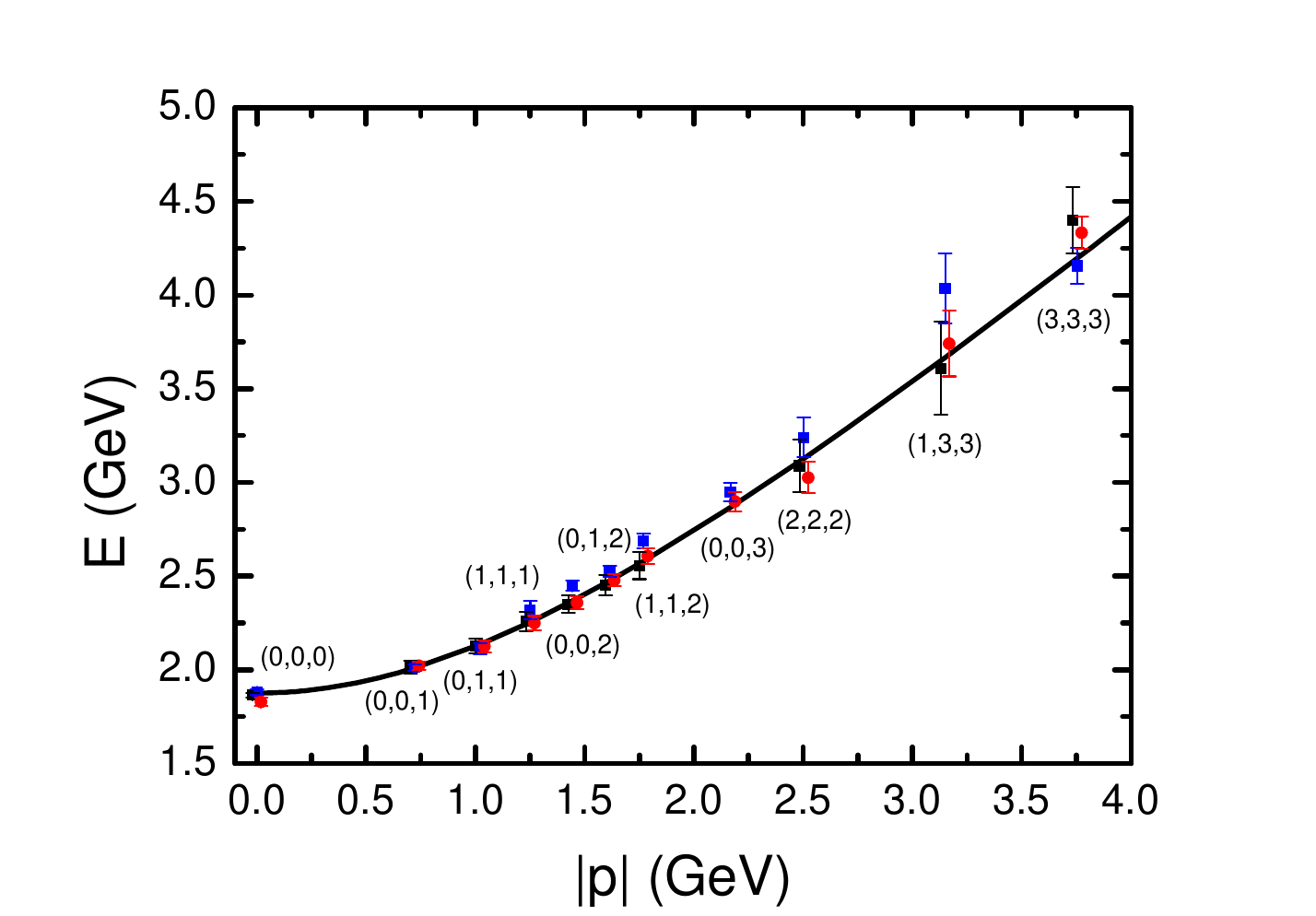}
\caption{ The fitted proton energies in physical units as a function
  of momentum on the $24^3\times 48$ lattice.
  The continuum dispersion relation is indicated by the black solid
  line.
  The symbols are the same as for Fig.~\ref{fg:correlationL24}, at the
  same eight momenta along with some additional $\vec{P}$ values.}
\label{fg:EandPL24}
\end{center} 
\end{figure}

The results of our fits as a function of momentum are shown in
Fig.~\ref{fg:EandPL24}, plotted against the continuum dispersion
relation.
In addition to the eight different proton momenta shown in
Figs.~\ref{fg:correlationL24}--\ref{fg:amfitL24}, we also add another
three momentum values, $P = (0,0,2)$, $ (0,1,2)$, $(0,0,3)$.
Note that, for the $P = (0,0,3)$ case, we are unable to find an
appropriate fit window in the case $N=1.$
The black solid line shows the continuum dispersion relation, $aE =
a\sqrt{m^2 + \vec{p}\,^2}.$
The use of the continuum dispersion relation (rather than a lattice
dispersion relation) is motivated by the fact that the nucleon is an
extended object, and relatively insensitive to physics at the scale of
a single lattice spacing.
We find that our lattice results are consistent with the continuum
dispersion relation at all of the momenta considered, and in
particular, the $N=8$ points (red circles) provide a very clean energy
dispersion all the way up to a proton momentum value of
$\vec{P}=(3,3,3)$.
  
\subsection{$32^{3}\times64$ results}
\label{sec:IIIc}
\begin{figure}[tbp]
\begin{center}
\includegraphics[width=0.45\textwidth]{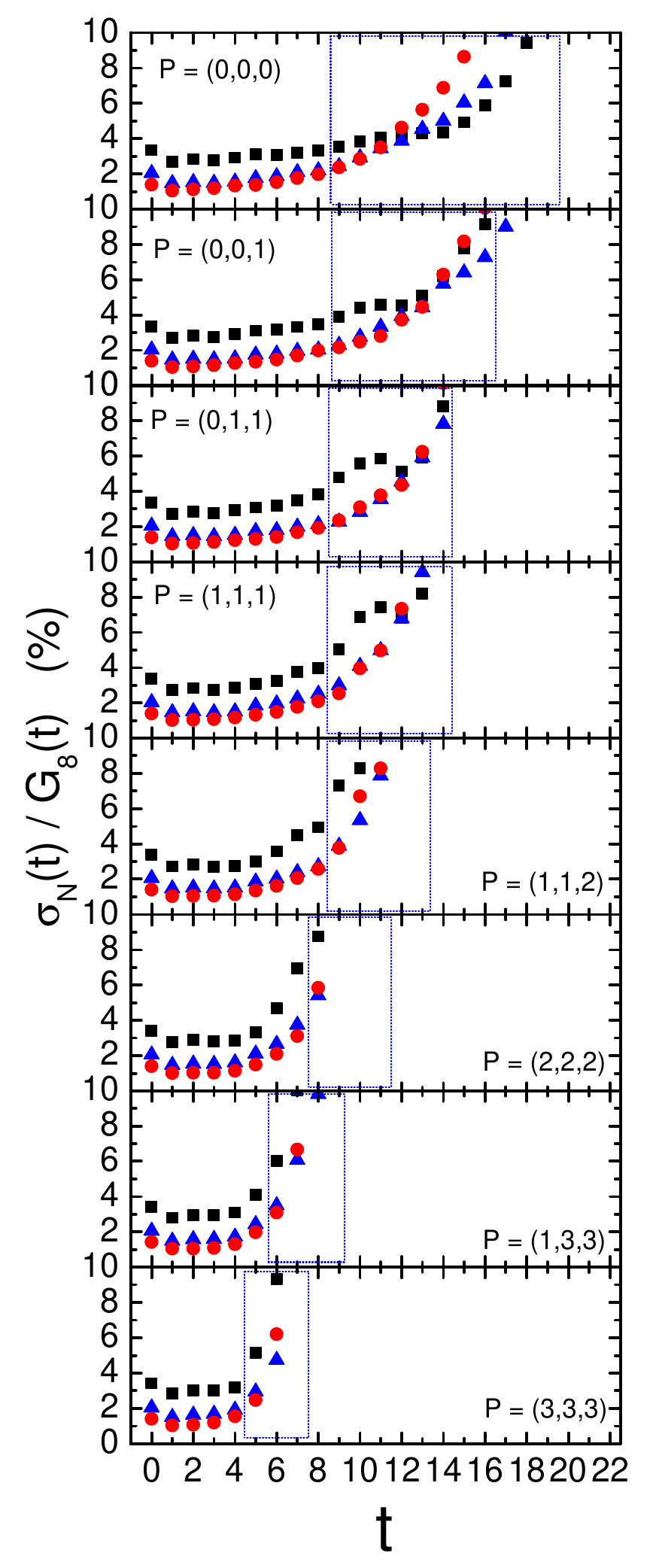}
\caption{The relative error in the proton correlation function with
  $N$ source points for the $32^3\times 64$ lattice volume.
The black square, blue triangle, and red circle points are for $N=1$,
$4$ and $8$ source locations, respectively.
The dashed blue boxes show the selected window for fitting the
effective energy.
}\label{fg:correlationL32}
\end{center} 
\end{figure}
\begin{figure}[tbp]
\begin{center}
\includegraphics[width=0.45\textwidth]{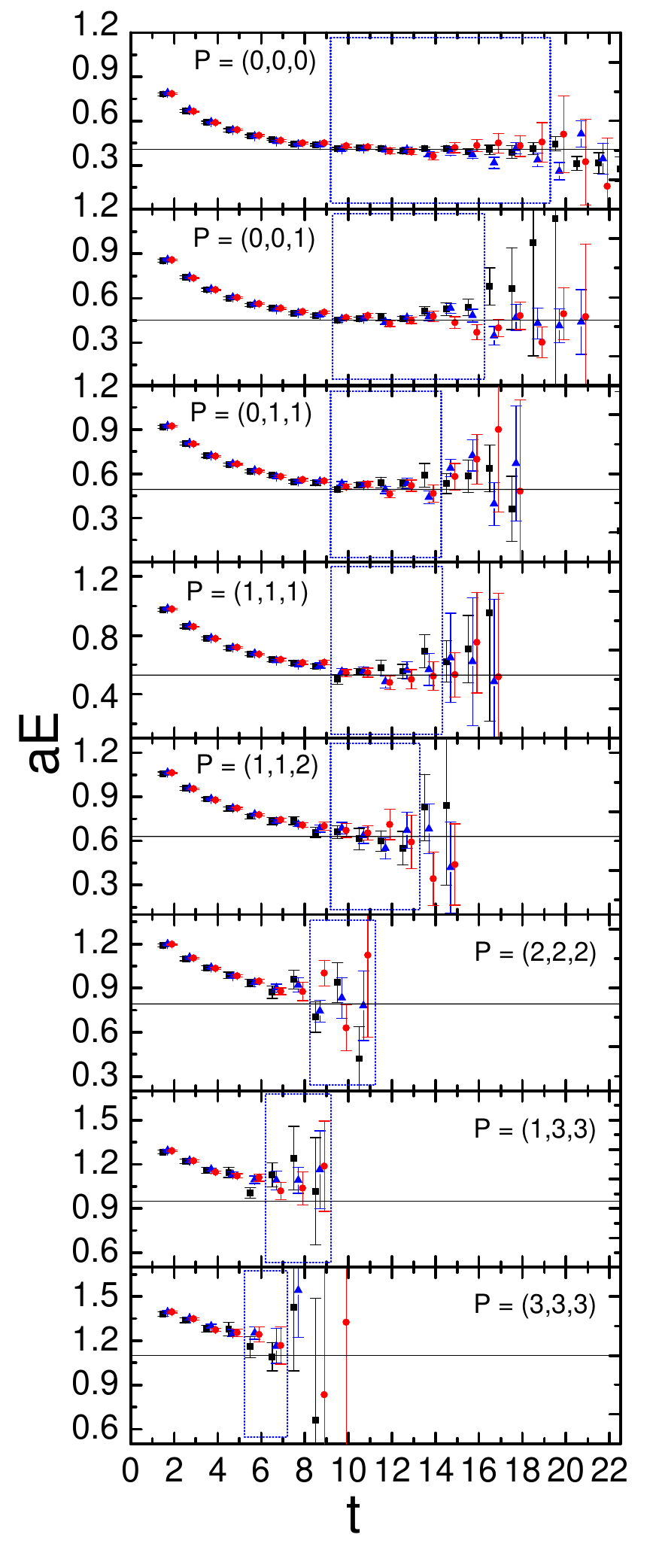}
\caption{
The proton effective energy in lattice units on the $32^3\times 64$
volume, with an inverse lattice spacing of $a=2.76\text{ GeV}^{-1}.$
Values are shown at eight momenta for $N=1,4,8$ with the same symbols
as in Fig.~\ref{fg:correlationL32}.
The solid line in the rest frame is the effective mass of proton,
$am$, calculated from the average of the three fitted effective masses
of the proton from the values of $N$.
The solid lines at nonzero momenta are calculated from the continuum
dispersion relationship, $ aE = a\sqrt{m^2 + \vec{p}\,^2}$.
The blue dashed boxes show the fitting range. 
}\label{fg:amdataL32}
\end{center} 
\end{figure}

%
We repeat the above investigation of the proton correlation function
on the $32^{3}\times64$ lattice volume, using 1000 configurations.
This ensemble is chosen as it has the same physical parameters as that
used in Ref.~\cite{Bali:2016lva}.

The normalised error $\sigma_N / G_8$
of the proton correlation
function with $N =1$, $4$ and $8$ source points is shown in
Fig.~\ref{fg:correlationL32}, with similar results to $24^3 \time 48$
lattice volume.
Again we see that (with the exception of the rest frame) the
correlation functions obtained using multiple source points have
smaller errors than for a single source point at all nontrivial
momenta.
The large time behaviour of the errors is also consistent with the
previous results, increasing more rapidly with higher values of $N.$

The proton effective energies at the eight different momenta
considered for the $32^3\times 64$ lattice are shown in
Fig.~\ref{fg:amdataL32}.
Up to a momentum value of $\vec{P} = (1,1,2)$, the plateaus obtained
are very clean, and we are able to fit starting from $t=9,$ with our
fit windows ending at $t \ge 13.$
At the three highest momenta values considered $\vec{P} = (2,2,2),$
$(1,3,3)$ and $(3,3,3),$ finding a suitable fit window is more
challenging.
The signal at higher momenta forces the fits to be made at earlier
times, and hence increases the risk of fitting in a region where true
single state dominance has not been achieved.

On the larger $32^3$ spatial volume, the momentum quantum $2\pi/L
\simeq 0.54$ GeV is smaller than for the $24^3$ lattice, so the
spacing between the different momentum states is reduced, potentially
increasing the amount of Euclidean time needed to achieve single state
dominance.
Though we have chosen not to perform a variational analysis in this
proof of concept study for reasons of simplicity, it is clear that the
use of correlation matrix techniques~\cite{Michael:1985ne,
  Luscher:1990ck, Blossier:2009kd, Mahbub:2009nr, Engel:2010my,
  Edwards:2011jj, Mahbub:2010rm, Mahbub:2013ala, Owen:2012ts,
  Kiratidis:2015vpa, Lang:2012db, Lang:2016hnn}, which remove excited
state contaminations and hence allow for fitting at earlier times,
will provide a significant advantage by leveraging the reduced
statistical errors offered by the use of a dilute grid source.
The problem of excited state contributions is further exacerbated at
higher momenta by the increasing presence of cross-parity
contaminations, however this may be controlled through the use of the
Parity-Expanded Variational Analysis (PEVA)
technique~\cite{Menadue:2013kfi}.
\begin{figure}[tbp]
\begin{center}
\includegraphics[width=0.5\textwidth]{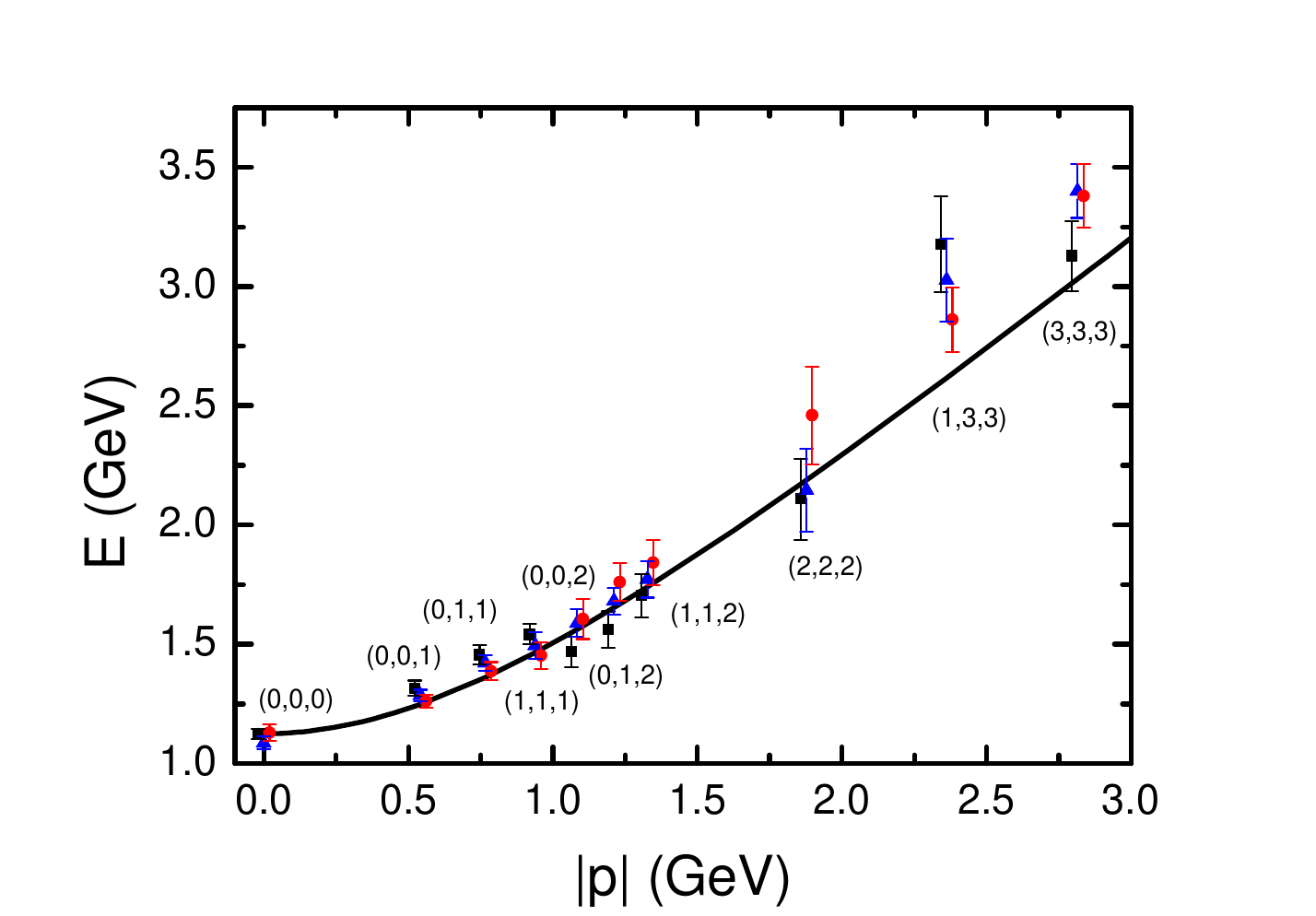}
\caption{ 
 The fitted proton energies in physical units as a function of
 momentum on the $32^3\times 64$ lattice.
 The continuum dispersion relation is indicated by the black solid
 line.
 The symbols are the same as for Fig.~\ref{fg:correlationL32}, at the
 same eight momenta along with some additional $\vec{P}$ values.}
\label{fg:EandPL32}
\end{center} 
\end{figure}

We can see potential hints of excited state contamination when we
compare the fitted proton energy results to the continuum dispersion
relationship in Fig.~\ref{fg:EandPL32}.
The agreement between the lattice results and the dispersion relation
is good up to $\vec{P} = (1,1,2),$ where we are fitting nice
plateaus.
At momentum values greater than this, starting with $\vec{P} =
(2,2,2),$ we begin to see some points that differ from the dispersion
relation at the level of $1\sigma-2\sigma.$
Noting that the points which disagree lie above the dispersion
relation, we infer that this is most likely to be an indication that
single state dominance has not been achieved before the signal is
overcome by noise, again emphasising the value in performing a
variational analysis (even for obtaining the lowest lying energy
state).
Nonetheless, in the selected fitting windows across all the non-zero
momentum results, we find that using $N=4$ source points provides
reduced statistical errors when compared with a single source point.
The quality of the $N=8$ results are generally as good or better than
$N=1$ within the fitting windows, but for certain momenta we find that
is not the case due to the rate at which the errors grow at later
times increasing with large $N.$

\section{Summary} \label{sec:summary}

%
We have introduced a novel stochastic source using $\mathbb{Z}_3$
noise placed on a dilute grid of lattice points, incorporating
iterative momentum-smearing.
The corresponding correlation function can be decomposed into two
parts, one incorporating the desired signal terms and the other
representing the noise terms that arise from the use of a stochastic
source.
The signal terms are the summation of various single point source
correlation functions, which can provide a more accurate signal than a
single point source for a fixed cost.
The benefit of this additional averaging competes with the additional
noise generated by the stochastic estimate of the double delta
function in the baryon one-end trick.

Through our numerical investigation, we find that there are two ways
to reduce the statistical uncertainties that arise from the noise
terms.
The first is to maximize the separation of the selected source points,
taking advantage of the quark propagator suppression increasing with
the distance.
The second is to choose an optimal number of source points, since the
number of noise terms increases much faster with $N$ than the number
of signal terms.
%

%
We performed calculations of the proton correlation function at a
variety of momenta on two lattice volumes, $24^3\times 48$ and
$32^3\times 64,$ using three different dilute grids with $N=1,4,$ and
$8$ maximally separarated source points.
Iterative momentum-smearing~\cite{Bali:2016lva} is applied at the
source and the sink to improve overlap with the boosted nucleon.
Using a dilute noise source, we can obtain an acceptable signal for
values of the total proton momentum up to $\vec{P}=(3,3,3).$
Our results show good agreement with the continuum dispersion relation
for the nucleon.

We find that for boosted systems, using multiple source locations
provides a better signal than that using a single point source, with
reduced statistical errors at early to mid Euclidean times.
At late times we find that the rate at which the error increases grows
with the number of source points $N$.
On the $24^3\times 48$ volume, the $N=4$ results are favoured at low
momenta within the selected fit window, with $N=8$ providing slightly
smaller errors at high momenta, though in general the differences
between using 4 and 8 source points are small.
The results with $N=4$ are favoured over all non-trivial momenta
considered on the $32^3\times 64$ volume.

In this proof of concept investigation, for simplicity we have not
used any correlation matrix techniques, instead comparing the
statistical errors obtained directly from the proton correlation
function.
Given that the statistical error reductions gained by using a dilute
grid source are greater at earlier times, there is a clear advantage
to be gained in using a variational method to eliminate excited state
contamination and hasten the onset of the single state dominance
region required for fitting.
Future work will incorporate the Parity Expanded Variational
Analysis~\cite{Menadue:2013kfi} technique to resolve the cross-parity
contributions at finite $\vec{p}$ in an effort to maximise the
advantages of the dilute noise source technique introduced here,
further enhancing our capability to study lattice hadrons in boosted
frames at high momentum values.

\acknowledgments

This work was supported by supercomputing resources provided by the
Phoenix HPC service at the University of Adelaide and the assistance
of resources from the National Computational Infrastructure (NCI).
NCI resources were provided through the National Computational Merit
Allocation Scheme, supported by the Australian Government and the
University of Adelaide Partner Share.  This research is supported by
the Australian Research Council through Grants No.~FT100100005,
CE110001004, FT120100821, DP140103067, DP150103164 and LE160100051.
  
\bibliography{boost}

\end{document}